\documentclass[10pt,twocolumn,prl,aps,amssymb,amsmath,tightenlines,showpacs]{revtex4}

\usepackage{graphicx}

\newcommand{\cP}{{\cal P}}
\newcommand{\cT}{{\cal T}}

\begin{document}

\title{Complex Correspondence Principle}
\author{Carl~M.~Bender${}^1$}\email{cmb@wustl.edu}
\author{Daniel~W.~Hook${}^2$}\email{d.hook@imperial.ac.uk}
\author{Peter~N.~Meisinger${}^1$}\email{pnm@physics.wustl.edu}
\author{Qing-hai~Wang${}^3$}\email{phywq@nus.edu.sg}

\affiliation{${}^1$Department of Physics, Washington University, St. Louis, MO
63130, USA \\
${}^2$Theoretical Physics, Imperial College London, London SW7 2AZ, UK\\
${}^3$Department of Physics, National University of Singapore, Singapore 117542}

\date{\today}

\begin{abstract}
Quantum mechanics and classical mechanics are two very different theories, but
the correspondence principle states that quantum particles behave classically
in the limit of high quantum number. In recent years much research has been done
on extending both quantum mechanics and classical mechanics into the complex
domain. This letter shows that these complex extensions continue to exhibit a
correspondence, and that this correspondence becomes more pronounced in the
complex domain. The association between complex quantum mechanics and complex
classical mechanics is subtle and demonstrating this relationship requires the
use of asymptotics beyond all orders.
\end{abstract}


\maketitle

\begin{figure}
\begin{center}
\includegraphics[scale=0.20, bb=0 0 1000 354]{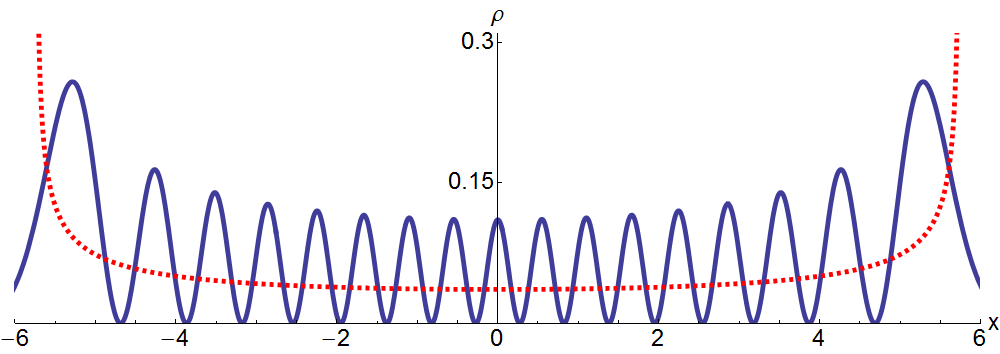}
\end{center}
\caption{Correspondence principle for the harmonic oscillator. The probability
density $\rho(x)=|\psi_{16}(x)|^2$ for a quantum particle (solid curve) and the
corresponding probability density for a classical particle of the same energy
(dotted curve) are shown. In the parabolic potential well $\rho_{\rm quantum}
(x)$ is wavelike and oscillates closely about $\rho_{\rm classical}(x)$. At the
classical turning points, $\rho_{\rm classical}(x)$ diverges. In the classically
forbidden region the quantum particle density decays exponentially while the
classical density is ordinarily assumed to vanish.}
\label{F1}
\end{figure}

The correspondence principle states that at high energy quantum mechanics
resembles classical mechanics. In Fig.~\ref{F1} we illustrate this resemblance
for the harmonic oscillator Hamiltonian $H=p^2+x^2$. We compare the quantum
probability density $\rho_{\rm quantum}(x)=|\psi_{16}(x)|^2$ for the 16th energy
level $E_{16}= 33$ with the classical probability density $\rho_{\rm classical}
(x)=1/s$, where $s$ is the speed of the particle. In the classically allowed
region, which is bounded by the classical turning points at $\pm\sqrt{33}$,
$\rho_{\rm quantum}(x)$ oscillates about $\rho_{\rm classical}(x)$. The particle
speed vanishes at the turning points, so $\rho_{\rm classical}$ is singular.
(This singularity is integrable and thus the classical probability is
normalizable.) In the classically forbidden region $\rho_{\rm quantum}(x)$
decays exponentially; in conventional classical mechanics the particle does not
enter this region and its probability density is thought to vanish.

\begin{figure}
\begin{center}
\includegraphics[scale=0.20, bb=0 0 1000 835]{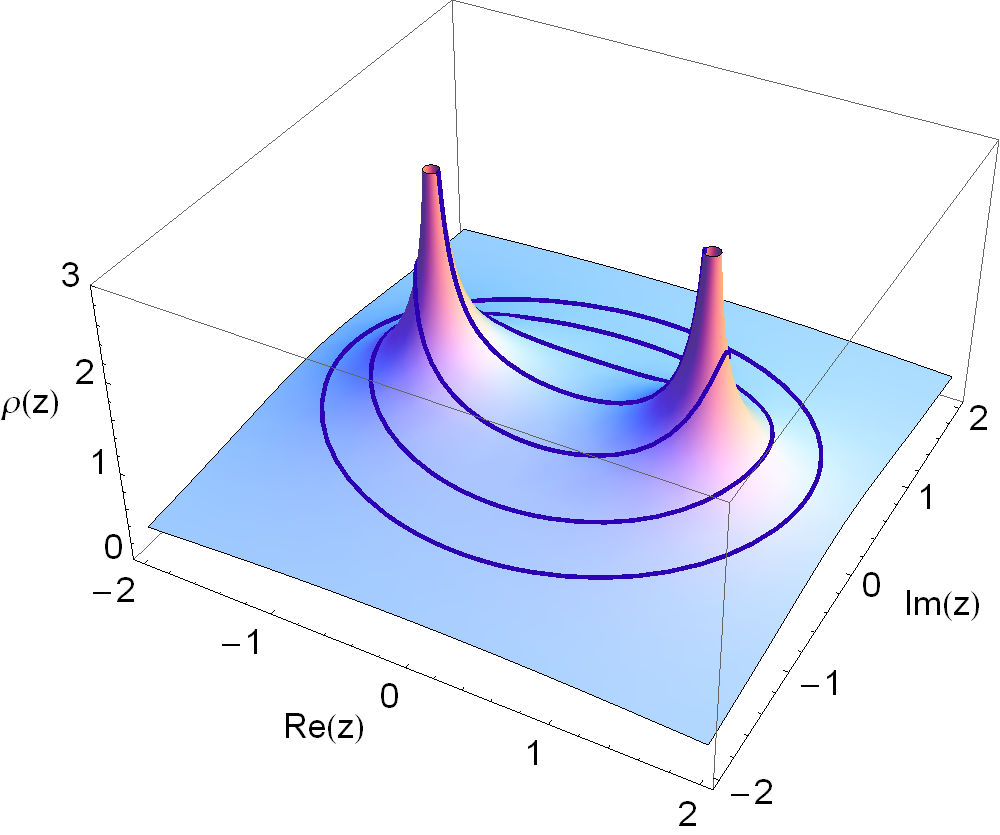}
\end{center}
\caption{Classical probability density at $z=x+iy$ in the complex plane for a
particle of $E=2$ in a harmonic potential. The probability density resembles a
pup tent with infinitely high tent poles located at the turning points. The
classical trajectories are nested ellipses, all of which have the same period
$T=\pi$. These complex orbits are superposed on the tent canopy. The degenerate
ellipse, whose foci are the turning points at $x=\pm\sqrt{E}$, is the
conventional real solution. Classical particles repeatedly cross the real axis
in the classically forbidden regions $|x|>\sqrt{E}$.} 
\label{F2}
\end{figure}

This paper generalizes quantum and classical probability into the complex
domain. In complex classical mechanics we solve Hamilton's equations ${\dot x}=
\frac{\partial H}{\partial p}$, ${\dot p}=-\frac{\partial H}{\partial x}$ for
complex initial conditions and not just for initial conditions in the
classically allowed region \cite{r1,r2,r3,r4,r5,r6,r7}. For the harmonic
oscillator the classical orbits are nested ellipses with foci at the turning
points. Using $\rho_{\rm quantum}$ in Fig.~\ref{F1} as a guide, we require the
probability of a classical particle being on more distant ellipses to decay
exponentially, and we plot in Fig.~\ref{F2} the relative probability density of
finding the classical particle at the point $z=x+iy$.

Figure \ref{F2} shows that, contrary to the traditional view, the classical
particle density in the classically forbidden region is actually nonzero, and
that $\rho_{\rm classical}$ resembles $\rho_{\rm quantum}$ in this region
\cite{r8}. This figure emphasizes that the classical probability density extends
beyond the real axis and into the complex plane, where $\rho_{\rm classical}(x,y
)$ falls off as the reciprocal of the distance from the origin. 

To establish the complex correspondence principle we must continue the quantum
probability density into the complex plane to match the complex classical
probability density in Fig.~\ref{F2}. To do so we must extend quantum mechanics
into the complex domain \cite{r9}. A Hermitian Hamiltonian ($H=H^\dagger$) has
real energy levels and unitary time evolution. However, the class of physically
allowable Hamiltonians may be extended to include non-Hermitian Hamiltonians
that possess an unbroken $\cP\cT$ (combined parity and time-reversal) symmetry
\cite{r10,r11,r12,r13,r14,r15}. These complex Hamiltonians also have real
spectra and generate unitary time evolution, and such Hamiltonians have recently
been observed in the laboratory \cite{r16,r17,r18}.

For $\cP\cT$-symmetric Hamiltonians the potential satisfies $V^*(-z)=V(z)$.
This allows us to derive a conservation law in the complex plane from the
time-dependent Schr\"odinger equation $i\psi_t(z,t)=-\psi_{zz}(z,t)+V(z)\psi(z,
t)$. The continuity equation is $\rho_t(z,t)+j_z(z,t)=0$, where the local
probability density and current are
\begin{eqnarray}
\rho(z,t)&\equiv&\psi^*(-z,t)\psi(z,t),
\label{e1}\\
j(z,t)&\equiv&i\psi_z^*(-z,t)\psi(z,t)-i\psi^*(-z,t)\psi_z(z,t).
\label{e2}
\end{eqnarray}
In this paper we limit our discussion to wave functions $\psi$ that are
eigenstates of $H$; for such states the density $\rho(z)$ is time independent
and the current $j(z,t)$ vanishes. The density $\rho(z)$ is {\it not} the
absolute square of $\psi(z,t)$ and it is {\it analytic} in the complex-$z$
plane.

A locally conserved probability density must be real and positive and its
spatial integral must be normalized to unity. While $\rho(z)$ satisfies a
continuity equation, it cannot be a probability density because it is
complex-valued. We propose a novel approach. We construct a complex contour $C$
satisfying three conditions:
\begin{eqnarray}
{\it Condition~I:}&\quad&{\rm Im}\,[\rho(z)\,dz]=0,\label{e3}\\
{\it Condition~II:}&\quad&{\rm Re}\,[\rho(z)\,dz]>0,\label{e4}\\
{\it Condition~III:}&\quad&\int_C{\rm Re}\,[\rho(z)\,dz]=1.\label{e5}
\end{eqnarray}

For brevity, we discuss the harmonic-oscillator ground state $\psi_0(z)=e^{-z^2/
2}$ for which $\rho(z)=e^{-x^2+y^2-2ixy}$. Since $dz=dx+idy$, Condition I gives
\begin{equation}
dy/dx=\sin(2xy)/\cos(2xy),
\label{e6}
\end{equation}
which is a differential equation for the contour $C$. On the contour defined by
this differential equation the local contribution $\rho(z)\,dz$ to the
probability is real.

Next, we turn to Condition III. Inserting (\ref{e6}) into (\ref{e5}), we get two
forms for the probability integral:
\begin{equation}
I=\int_C dx\frac{e^{[y(x)]^2-x^2}}{\cos[2y(x)x]}
=\int_C dy\frac{e^{y^2-[x(y)]^2}}{\sin[2yx(y)]}.
\label{e7}
\end{equation}
These integrals converge if the contour $C$ terminates in the two {\it good}
Stokes' wedges of opening angle $\pi/2$ centered about the real axis, but they
diverge if $C$ terminates in the {\it bad} Stokes' wedges of opening angle $\pi/
2$ centered about the imaginary axis.

To determine whether the contour $C$ terminates in good Stokes' wedges we must
solve (\ref{e6}). However, this simple-looking differential equation does not
possess a closed-form solution, and we must rely on asymptotic techniques. For
large $x$ the contour $C$ approaches the center of the good Stokes' wedge
and for large $y$, $C$ approaches the center of the bad Stokes' wedge:
\begin{eqnarray}
y(x)&\sim&n\pi/(2x)\quad(x\to+\infty,\,n\in\mathbb{Z}),
\label{e8}\\
x(y)&\sim&(m+1/2)\pi/(2y)\quad(y\to+\infty,\,m\in\mathbb{Z}).
\label{e9}
\end{eqnarray}

\begin{figure}
\begin{center}
\includegraphics[scale=0.19, bb=0 0 1000 925]{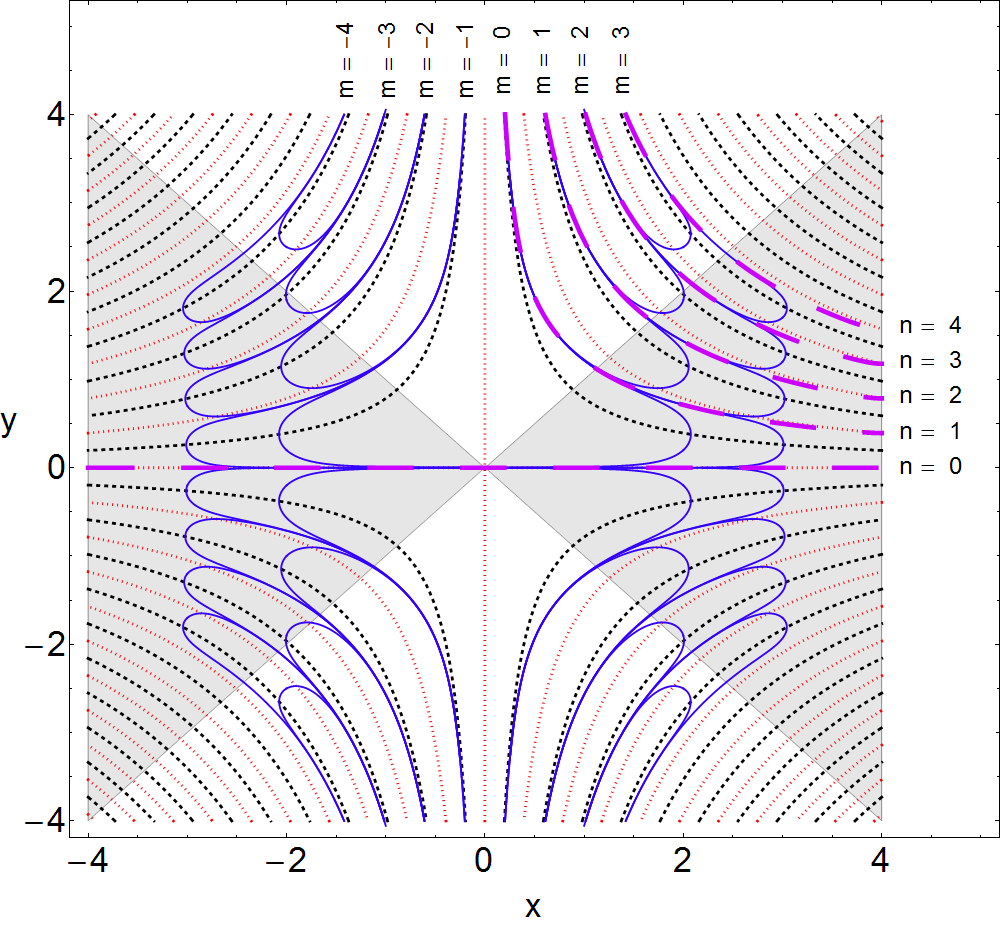}
\end{center}
\caption{Numerical solutions (solid curves) to the differential equation
(\ref{e6}) in the complex-$z=x+iy$ plane. Solutions have vanishing slope on
lightly dotted hyperbolas and infinite slope on heavily dotted hyperbolas. In
the {\it good} Stokes' wedges (dark shading) $\rho(z)$ decays exponentially and
in the {\it bad} Stokes' wedges (unshaded) $\rho(z)$ grows exponentially. The
only curves that reach $\pm\infty$ in the good wedges are separatrices (heavy
dashed curves labeled $n=0,\,\ldots,\,4$). A typical solution curve in a good
Stokes' wedge is unstable; as $x\to\infty$ the curve turns away from the
separatrix, leaves the wedge, and is drawn into a bad Stokes' wedge. In the bad
wedge curves are stable and continue on to $\pm i\infty$ in quantized bunches
labeled by $m$. The only continuous unbroken curve connecting the two good
wedges is the special separatrix on the real axis.}
\label{F3}
\end{figure}

This asymptotic analysis shows that the integration paths occur in {\it
quantized} bunches labeled by the integers $m$ or $n$. Higher-order asymptotic
analysis reveals that these bunches are {\it stable} in the bad Stokes'
wedges and {\it unstable} in the good Stokes' wedges. This means that as
$|y|\to\infty$, a path in the bad Stokes' wedge is drawn towards the quantized
curves in (\ref{e9}), but that as $|x|\to\infty$ paths veer away from the
quantized curves in (\ref{e8}), which we call {\it separatrices}. Thus, only the
isolated separatrices ever reach $\infty$ in the good Stokes' wedges; all other
curves turn around and are drawn into the bad Stokes' wedges. This remarkable
behavior is illustrated in Fig.~\ref{F3}.

Figure~\ref{F3} shows that the only continuous path connecting the two good
Stokes' wedges follows the real axis. All other paths terminate in one good and
one bad Stokes' wedge, and thus the probability integral (\ref{e7}) along
such paths diverges. This instability of contours in the good Stokes' wedge in
Fig.~\ref{F3} is a serious and generic problem that must be overcome if we wish
to extend the quantum probability density into the complex plane.

To overcome this apparently insurmountable problem, we show that if a contour
satisfying (\ref{e6}) in the complex-$z$ plane enters a bad Stokes' wedge along
one path and then returns along a second path {\it in the same quantized
bundle}, then the integral along the combined paths is actually convergent! This
result is surprising because the integral along each path separately is
exponentially divergent. To prove convergence, we examine the integral
\begin{equation}
I=\int_Y^\infty dy\,e^{y^2}\left(\frac{e^{-[u(y)]^2}}{\sin[2yu(y)]}-
\frac{e^{-[v(y)]^2}}{\sin[2yv(y)]}\right),
\label{e10}
\end{equation}
where $u(y)$ and $v(y)$ are two solutions to (\ref{e6}) in the same bunch (that
is, having the same value of $m$). The leading asymptotic behaviors of these
solutions are given in (\ref{e9}), and in fact the entire Poincar\'e asymptotic
expansions of $u(y)$ and $v(y)$ are identical to all orders in powers of $1/y$.
However, the difference $D(y)\equiv u(y)-v(y)$ is nonzero and is exponentially
small
\begin{equation}
D(y)\sim Ce^{-y^2}\left[1+(n+1/2)^2\pi^2y^{-2}/4+\ldots\right]
\label{e11}
\end{equation}
as $|y|\to\infty$, where $C$ is an arbitrary constant. This result reflects the
{\it hyperasymptotic} (asymptotics beyond all orders) content of (\ref{e6})
\cite{r19}.

\begin{figure}
\begin{center}
\includegraphics[scale=0.21, bb=0 0 1000 467]{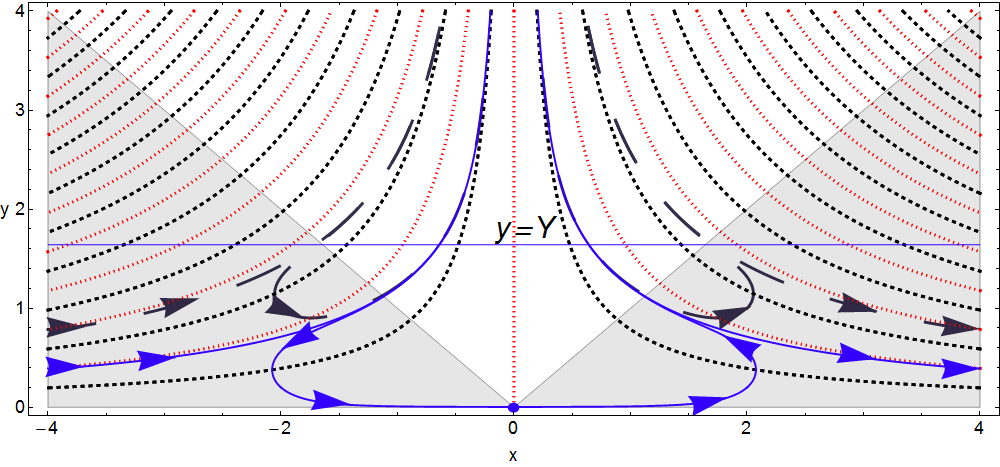}
\end{center}
\caption{Complex contours connecting good Stokes' wedges. A contour (solid
curve) begins at $x=-\infty$ in the left good Stokes' wedge (shaded), leaves the
wedge along a separatrix, and runs up to $i\infty$ in the bad Stokes' wedge
(unshaded). The contour continues downward along a path in the same bunch,
crosses the imaginary axis at $y=0.003\,324\,973\,872\,707\,912$, and heads
upwards into the same bad Stokes' wedge. Finally, the contour re-emerges from
the bad Stokes' wedge and continues towards $x=+\infty$ along a separatrix in
the right good Stokes' wedge. Contours have zero slope on lightly dotted lines
and infinite slope on heavily dotted lines. A second contour (dashed curve)
connecting the two good Stokes' wedges visits the bad Stokes' wedge four times
instead of twice.}
\label{F4}
\end{figure}

Using (\ref{e11}), we approximate the integral (\ref{e10}):
\begin{equation}
I\sim C(-1)^n(n+1/2)\int_Y^\infty dy\,y^{-3}\quad(Y\to\infty),
\label{e12}
\end{equation}
which is convergent. Thus, while there is no continuous path running between the
two good Stokes' wedges, there do exist paths connecting these wedges that
repeatedly enter and re-emerge from bad Stokes' wedges. On these contours the
probability integral $I$ in (\ref{e7}) exists. Several such paths are shown in
Fig.~\ref{F4}.

\begin{figure}
\begin{center}
\includegraphics[scale=0.20, bb=0 0 1000 1018]{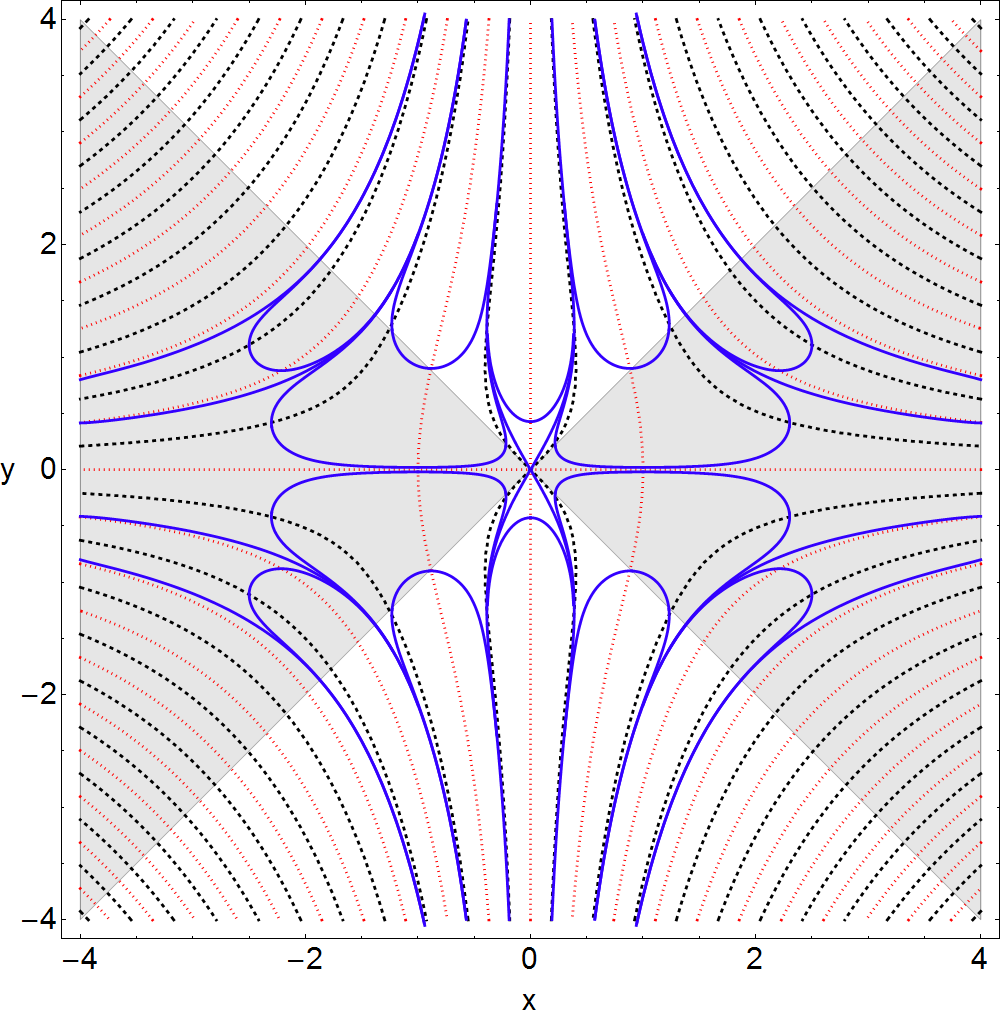}
\end{center}
\caption{Complex contours for the first excited state of the quantum harmonic
oscillator. Four contours (solid lines) that begin at $x=-\infty$ in the left
good Stokes' wedge (shaded) are shown. These contours leave this wedge along
separatrices and run off to $\pm i\infty$ in the upper and lower bad Stokes'
wedges (unshaded). After visiting a bad Stokes' wedge several times, the
contours may pass through the node at the origin at $60^\circ$ angles to the
horizontal. At this node the probability density vanishes. Then the contours
repeat the process in the right-half plane and eventually enter the right good
Stokes' wedge along separatrices. It is also possible to have a complex contour
that avoids the node at the origin and still connects the good Stokes' wedges.
Solution curves are horizontal on lightly dotted lines and vertical on heavily
dotted lines.}
\label{F5}
\end{figure}

Figure \ref{F5} is a generalization of Fig.~\ref{F4} for the first excited state
of the quantum harmonic oscillator. Note that contours that connect the two
good Stokes' wedges may or may not pass through the node at the origin.

We do not present the argument in this paper, but along the contours shown in
Figs.~\ref{F4} and \ref{F5} Condition II in (\ref{e4}) is satisfied. That is,
along the entire complex contour obeying the differential equation (\ref{e6})
the local contribution to the total probability integral is positive. Thus,
the requirements in the three conditions (\ref{e3} -- \ref{e5}) are met, and we
have successfully extended the probabilistic interpretation of quantum mechanics
into the complex plane.

\begin{figure}
\begin{center}
\includegraphics[scale=0.22, bb=0 0 1000 1014]{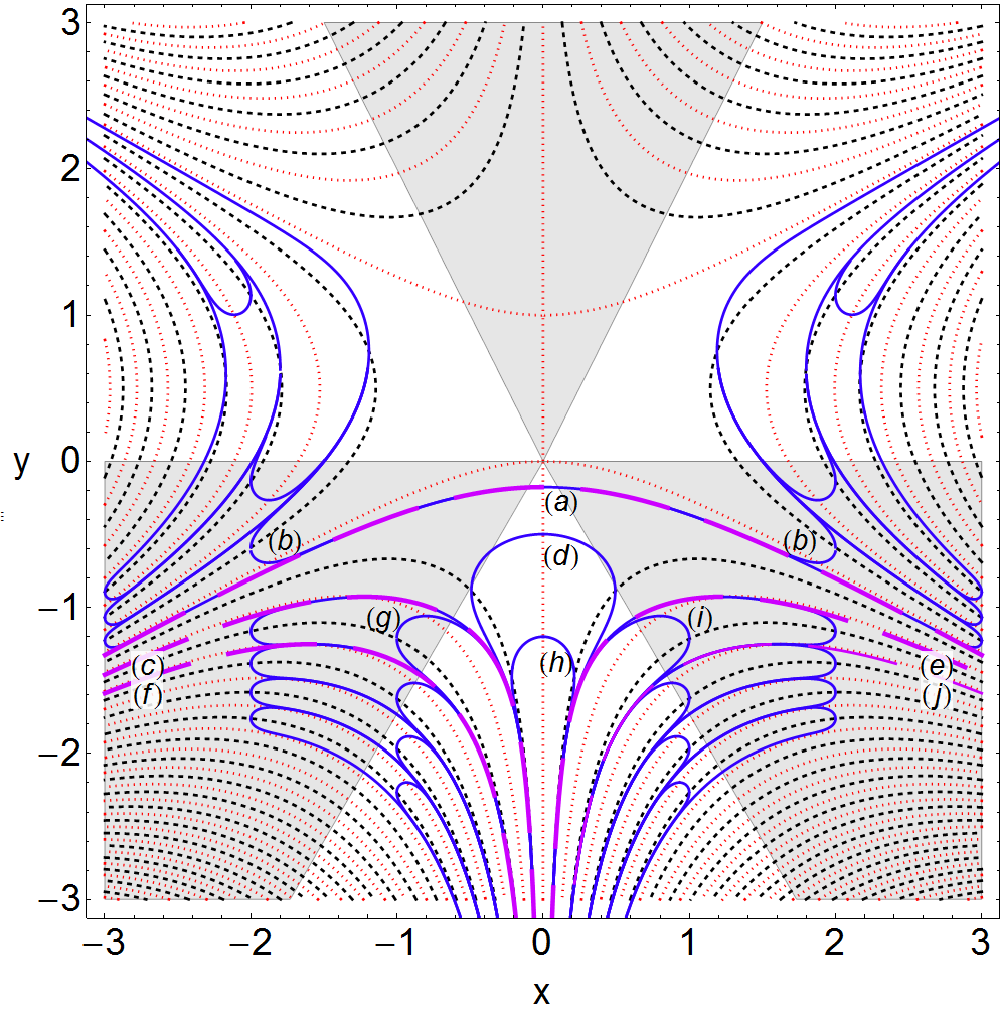}
\end{center}
\caption{Complex contours (heavy dashed lines) for the ground state ($J=1$) of
the quasi-exactly-solvable $\cP\cT$ anharmonic oscillator. A separatrix (a) goes
directly from the left good Stokes' wedge to the right good Stokes' wedge,
crossing the imaginary axis at $y=-0.176\,651\,795\,619\,462\,368$. Paths (b)
that cross the imaginary axis at higher points than the (a) path cannot reach
$\infty$ in the good wedge. Because they are unstable, such paths turn around,
enter the upper bad Stokes' wedges, and can never re-emerge. A separatrix (c)
leaves the left good wedge. This path enters the lower bad wedge to the left of
the imaginary axis, re-emerges along paths (d) or (h), and reenters the bad
wedge to the right of the imaginary axis. It then continues into the right good
wedge along the separatrix (e). Another separatrix (f) leaves the left good
Stokes' wedge, enters the lower bad Stokes' wedge, leaves and returns along (g),
leaves, reenters again along (d) or (h), leaves and reenters along (j), and
finally enters the right good wedge along the separatrix (f). Solution paths are
horizontal on lightly dotted lines and vertical on heavily dotted lines.}
\label{F6}
\end{figure}

The analysis in this paper is general and extends to other $\cP\cT$ quantum
theories, such as the quasi-exactly solvable quartic anharmonic oscillator,
whose Hamiltonian is $H=p^2-x^4+2ix^3+x^2+2iJx$ \cite{r20}. However, for quantum
theories other than the harmonic oscillator, there are two kinds of bad Stokes'
wedges: (1) wedges for which there exist probability contours $C$ that enter and
re-emerge from the wedge and for which the probability integral along $C$
converges and (2) wedges for which no such contour exists. For the Hamiltonian
above there are three good Stokes' wedges and three bad Stokes' wedges, one of
type (1) and two of type (2) (see Fig.~\ref{F6}).

The quantum probability densities associated with the curves in Figs.~\ref{F4},
\ref{F5}, and \ref{F6} resemble pup tents with ripples in their canopies, in
contrast to the smooth surface in the classical pup tent in Fig.~\ref{F2}. This
is the complex analog of Fig.~\ref{F1} and illustrates the complex
correspondence principle.

CMB thanks the U.S.~Department of Energy and DWH thanks Symplectic Ltd.~for
financial support.

\end{document}